\documentclass[preprint,showpacs,prd]{revtex4}
\usepackage{mathrsfs}
\usepackage{epsfig,amsmath}
\usepackage{graphicx}
\usepackage{bm}

\begin{document}
\title{\textbf{Influence of the intensity gradient upon HHG from free electrons scattered by an intense laser beam}}
 \author{Ankang Li, Jiaxiang Wang \footnote{{jxwang@phy.ecnu.edu.cn}}, Na Ren}

\address{State Key Laboratory of Precision Spectroscopy, East China Normal University, Shanghai 200062, China}

\author{Pingxiao Wang}
\address{Applied Ion Beam Physics Laboratory, Key Laboratory of the Ministry of Education ,
China and Institute of Modern Physics, Department of Nuclear Science and Technology,
Fudan University, Shanghai 200433, China}

\author{Wenjun Zhu, Xiaoya Li}
\address{National Key Laboratory of Shock Wave and Detonation Physics, Mianyang 621900, Sichuan, China}

\author{Ross Hoehn, Sabre Kais}
\address{Departments of Chemistry and Physics, Purdue University, West Lafayette, Indiana 47907, USA}
\address{Qatar Environment and Energy Research Institute, Qatar Foundation, Doha, Qatar}

\def\abstractname{}
\begin{abstract}
\hspace*{\fill}{\bf Abstract}\hspace*{\fill}\\
When an electron is scattered by a tightly-focused laser beam in vacuum, the intensity gradient is a critical factor to influence the electron dynamics,
for example, the electron energy exchange with the laser fields as have been explored before [P.X.Wang et al.,J. Appl. Phys. 91, 856 (2002]. In this paper, we have further investigated its influence upon the electron high-harmonic generation (HHG) by treating the spacial gradient of the laser intensity as a ponderomotive potential. Based upon perturbative QED calculations, it has been found that the main effect of the intensity gradient is the broadening of the originally line HHG spectra. A one-to-one relationship can be built between the beam width and the corresponding line width. Hence this finding may provides us a promising way to measure the beam width of intense lasers in experiments. In addition, for a laser pulse, we have also studied the different influences from transverse and longitudinal intensity gradients upon HHG.

\end{abstract}
\maketitle
\section*{\textbf{1. Introduction}}

Since the availability of high-power lasers, high harmonic
generation (HHG) based upon nonlinear Compton scattering (NLCS) from
free electrons in strong laser fields has drawn considerable
attention \cite{1,2,3,4,5,6,7,8,9,10}. This is not only because it
is a fundamental non-perturbative laser-induced phenomena, but also
because it is a prospective x-ray or gamma-ray source
\cite{11,12,13} with remarkable performances in terms of tunability.
Moreover, the free-electron laser interaction is very clean without
other uncontrollable physical processes such as ionizations and
collisions, which happen in the interaction between lasers and atoms
or plasmas. In recent years, the observation of NLCS in some
experiments \cite{14,15,16} also renew the interest in its
theoretical study.

The main aim of this paper is to investigate how the effect of the
laser intensity gradients change the radiation of free electron in
strong laser field. It is well known that there is a cycle-averaged
force on a charged particle in a spatially inhomogeneous laser
field. It is associated with a time-independent potential energy
called ponderomotive potential, which is due to the laser intensity
gradient in an oscillatory field. After the presence of such
ponderomotive potential was first proposed in 1957 by Boot and
Harvie \cite{17,18}, it is well known for the last four decades that
the potential could have a significant effect on the matter
interacting with the laser field, such as particle acceleration
\cite{19}, trapping and cooling of the atoms \cite{20} , high-field
photoionization of atoms \cite{21}, self-focusing in plasma
\cite{22} and high harmonic generation \cite{23,24}. In the
classical framework, the mechanical motion of electrons in a strong
laser field will be changed if the ponderomotive potential is taken
into account due to the limited spatial dimensions of the laser
focus, which leads to the ponderomotive broadening of the radiation
spectrum \cite{23}. But so far few works are done to study the role
of the ponderomotive effects on the radiation spectrum based on a
quantum theory \cite{23,24,31}.

To gain a clear idea of the influence by the laser intensity
gradients on the HHG spectrum from free electrons in strong laser
fields, we start from the scattering of the electron Volkov state by the ponderomotive potential of the laser beam. The corresponding cross section is
calculated as a second order quantum electrodynamics (QED)
laser-assisted process similar to laser-assisted bremsstrahlung,
where an charged particle scatters by the field of a nucleus in a
background strong laser field \cite{25,26,27,28,29,30}.

For notations in this paper, the four-vector product is denoted by {$a\cdot
b=a^0b^0-\bm{ab}$} and the Feynman dagger is {$\rlap{\slash}A=\gamma\cdot A$}. The Dirac adjoint is
denoted by {$\overline{u}=u^{\dagger}
\gamma^0$} for a bispinor u and {$\overline{F}=\gamma^0
F^{\dagger}\gamma^0$} for a matrix F.

The outline of this paper is the following. First, we will introduce
the scattering model and derive the theoretical expression for the
cross section of the electron radiation. Then, the numerical
estimation of the cross section and the corresponding analyses will
be provided in Sec 3. Concluding remarks are reserved for Sec 4.

\section*{2. Theoretical derivation of the scattering cross section}

We begin by introducing our scattering mode: Consider there are two
monochromatic laser pulse in space. The first one propagates along z
axis, of which the duration is long enough and the field intensity
so strong that it can be modeled by a background classical plane
wave field, described by a four dimension vector potential
{$A^{\mu}$}:

\begin{eqnarray}
{A^{\mu}}&=&A_0[{\delta}\cos\phi{\epsilon_1}^\mu+(1-\delta^2)^{1/2}\sin\phi{\epsilon_2}^\mu],\label{field}
\label{plane}
\end{eqnarray}

Here with the phase factor {$\phi=k\cdot x$}, in which $x$ is the
position vector, and the four wave vector is related to
{$k^\mu=\frac{\omega_0}{c}(1,0,0,1)$} with $\omega_0$ denoting the
wave propagation direction and laser frequency, respectively. The
laser is circularly polarized for $\delta=1/\sqrt{2}$ and linearly
polarized for $\delta=0,\pm1$. We define two polarization vectors
$\epsilon_1$, $\epsilon_2$, satisfying $\epsilon_i\cdot k=0,\,
\epsilon_i\cdot\epsilon_j =\delta_{ij}$ $(i,j=1,2)$. The laser
intensity can be easily described by a dimensionless parameter
$Q=eA_0/(mc^2)$, which is usually called laser intensity parameter.
In the nonrelativistic regime, the characteristic velocity and
energy for an electron moving in such an electromagnetic field is
$v\sim{eA_0/(mc)}$ and $E\sim{e^2A_0^2/(mc^2)}$, so relativistic
treatment is necessary if $v\sim c$ and $E\sim{mc^2}$ is satisfied.
That's to say the motion of the electron will become relativistic
for $Q\sim1$. In the optical regime ($\hbar\omega\approx1eV$), the
corresponding laser intensity is about $10^{18}W/cm^2$ for $Q\sim1$,
which has been achieved in the last decade.

\begin{figure}
\includegraphics[scale=0.4]{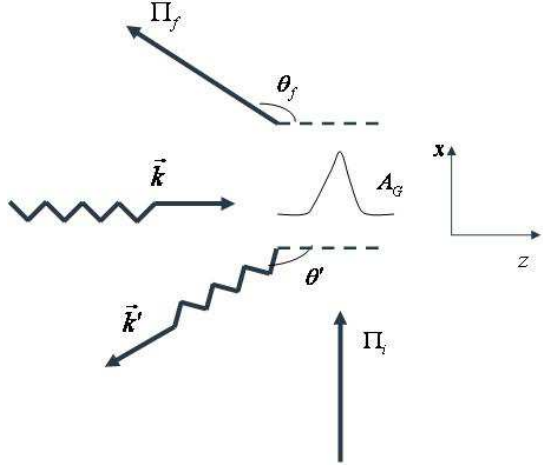}
\caption{\footnotesize{ The scattering geometry: The incoming
electron with four momentum $p_i$ have a $90^\circ$ collision with a
plane laser field while scattering by a time-independent
ponderomotive potential $A_G$. The final electron with $\Pi_f$ and
the emitted photon with $k^\prime$ are projected onto the xz plane
in this figure; $\theta_f$ and $\theta^{\prime}$ denotes the
scattering angle of the outgoing electron and the emitted
photon,respectively.}}
\end{figure}
%


The second field is a tightly focused laser pulse propagating
opposite the first one, which can be described by the lowest-order
axicon Gaussian fields with a envelop factor
{$g(\phi)=e^{\big(-{\phi^2}/{(\Delta\phi)}\big)}$}. It is far less
intense than the first one (the intensity is described by another
dimensionless parameter {$Q_G$})  but much rapidly oscillatory. The
electron will be assumed to be moving with a momentum p in the
direction perpendicular to z axis, say, x axis. When the electron is
placed in the two laser wave field, it will be mainly driven by the
longer-wavelength laser. The interaction of the electron with the
plane wave laser will be treated exactly by introducing the
well-known Volcov state. That's to say, the "dressed" electron will
have an effective momentum {$\Pi=p+\frac{e^2A_0^2}{4c^2(p\cdot k)}$}
with a corresponding effective mass
{$m^{\prime}=m\sqrt{1+\frac{Q^2}{2}}$}.While the main effect of the
Gaussian laser pulse on the electron can be described by an
time-averaged potential in view of its low intensity and fast
oscillation (here we do not care about the high frequency part of
the radiation caused by the fast oscillation). We assume that the
relativistic electron moves so fast that the envelop factor is
almost time independent. The effective pondermotive potential can be
described by :

\begin{eqnarray}
U_p=\frac{mc^2}{4}{Q_G}^2e^{-\frac{\bm{r_\perp}^2}{{b_0}^2}-\frac{z^2}{{b_1}^2}}
\end{eqnarray}

The corresponding four vector potential can be written as:
\begin{eqnarray}
A^{\mu}_p=\frac{U_p}{e}\delta^{\mu0}
\end{eqnarray}

where the Fourier-transformed pondermotive potential is given by:
\begin{eqnarray}
A^{\mu}_p(\bm{q})=\frac{mc^2\pi^{3/2}}{4e} {Q_G}^2 {b_0}^2 b_1
\delta^{\mu0}e^{-\frac{{b_0}^2}{4}\bm{q_\perp}^2-\frac{{b_1}^2}{4}{q_z^2}}
\end{eqnarray}

Here {$\bm{r_\perp}$}  and {$\bm{q_\perp}$}  refers to the position
and momentum in the plane perpendicular to the propagation dirction,
respectively. While is related to the beam waist and  is decided by
the pulse duration. Considering the high-frequency character of the
pulse, which means a small wavelength, the paraxial approximation of
the laser pulse can be acceptable confidently. Since $Q_G\ll Q$ , we
will treat the pondermotive potential as perturbation. The whole
scattering configuration can be seen in Fig 1 and it can be
described by two Feynman diagrams shown in Fig 2.

\begin{figure}
\includegraphics[scale=0.4]{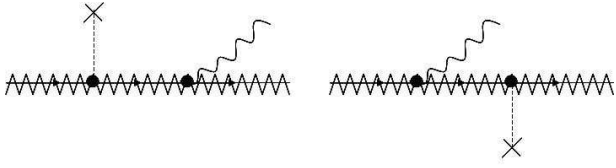}
\caption{\footnotesize{Feynman diagrams describing laser-assisted
bremsstrahlung. The laser-dressed electron and laser-dressed
electron propagator are denoted by a zigzag line on top of the
straight line. The Coulomb field photon is drawn as a dashed line,
and the bremsstrahlung photon as a wavy line.}}
\end{figure}
%
It's obvious that the interaction of electron with the intense plane
laser wave will leads to discrete line at a given harmonic on the
spectrum if the second weak pulse is absent. The scattering process
is called laser-induced Compton scattering, or Nonlinear Compton
scattering (NLCS) for its nonlinear nature, which has been widely
studied since the invention of laser in 1960 \cite{9,10,11}. The
frequencies of the emitted harmonics are determined from the
energy-momentum conservation laws, which involve both the incident
electron condition and laser parameters. Here we wonder the effect
of the pondermotive potential on the discrete radiation spectrum.

In order to calculate the differential cross-section of the second
order laser-assisted process, we begin by employing the well-known
Volkov state to describe the initial and final electron state:

\begin{eqnarray}
\psi_{p,r}=\frac{1}{V}\sqrt{\frac{mc}{\Pi^0}}\zeta_p(x)u_r(p) \\
\zeta_p(x)=\big(1+\frac{e\rlap{$\slash$}k\rlap{$\slash$}A}{2p\cdot
k}\big)e^{iS}
\end{eqnarray}

\begin{eqnarray}
S=&-&\frac{\Pi\cdot x}{\hbar}-\frac{e^2A_0^2}{8\hbar
c^2(p\cdot k)}(2\delta^2-1)\sin2\phi+\frac{eA_0}{\hbar c(p\cdot k)}\nonumber\\
&\times&\big[\delta({p}\cdot{\epsilon_1})sin\phi
-(1-\delta^2)^{1/2}({p}\cdot{\epsilon_2})\cos\phi\big].
\end{eqnarray}

Here $u_r(p)$ the free Dirac spinor. Here we employ a box
normalization with a normalization volume V.

Then the corresponding transition amplitude can be written as:
\begin{eqnarray}
S_{fi}=-\frac{e^2}{\hbar^2 c^2}{\int
dx^4dy^4\bar{\psi}_{p_f,r_f}(x)[\rlap{\slash}A_c(x)iG(x-y)\rlap{$\slash$}A_G(y)+\rlap{\slash}A_G(x)iG(x-y)\rlap{$\slash$}A_c(y)]
\psi_{p_{i},r_{i}}(y)}\nonumber \\
\end{eqnarray}

{$A_c^\mu(x)=\sqrt{\frac{2\pi\hbar}{\omega^\prime}}c{\epsilon_c}^\mu
e^{i{k^\prime} x}$} stands for the four-momentum of the emitted
photo during the scattering process with {${\epsilon_c}^\mu$} the
photon polarization vector and
{$k^\prime=\frac{\omega^\prime}{c}(1,\bm{e_{k^\prime}})$} the wave
vector, respectively. As the previous work \cite{28,29}, here we use
the Dirac-Volkov propagator instead of the free electron propagator
in view of the strong laser field:

\begin{eqnarray}
iG(x-y)=-\int \frac{dp^4}{(2\pi\hbar)^3(2\pi
i)}\zeta_p(x)\,\frac{\rlap{\slash}p+mc}{p^2-m^2c^2}\,\bar{\zeta}_p(y)
\end{eqnarray}

It has been proven\cite{28} that the use of the Dirac-Volkov
propagator is crucial to obtain correct numerical result in
laser-modified QED process.

Here we take average on the initial electron spin (i.e, the electron
is unpolarized) and sum over the both final electron spin and
emitted photon polarization. The differential cross section is
calculated with the formula:

\begin{eqnarray}
d\stackrel{\sim}{\sigma}=\frac{1}{2JT}\sum_{r_i,r_f,\varepsilon_c}
\big|S_{fi}\big|^2\frac{Vd^3\Pi_f}{(2\pi\hbar)^3}\frac{d^3k^\prime}{(2\pi)^3}.
\end{eqnarray}

Here T is the long observation time and
{$J=\frac{c}{V}\frac{\bm{\Pi}}{\Pi^0}$} stands for the incoming
particle flux. We have
$d^3\Pi_f=|\bm{\Pi_f}|^2d\Omega_f=|\bm{\Pi_f}|^2sin\theta_fd\theta_fd\varphi_f$,
$d^3k^{\prime}=\frac{{\omega^{\prime}}^2}{c^2}d\Omega^{\prime}=\frac{{\omega^{\prime}}^2}{c^2}sin\theta^{\prime}d\theta^{\prime}d\varphi^{\prime}$,where
$\Omega^{\prime}$ and $\Omega_f$ stands for the the solid angle of
the emitted photon and electron, respectively. After a long but
straight-forward deriving process, we finally write down the
differential cross-section as:

\begin{eqnarray}
\frac{d\stackrel{\sim}{\sigma}}{d\omega^\prime d\Omega^\prime
d\Omega_f}&=&\frac{\alpha{Q_G^4}{b_0^4}}{8(4\pi)^4c^2}(\frac{m^2c^2}{\hbar^2}b_1^2)\nonumber\\
&&\times\;\sum_{n,\varepsilon_c}\frac{|\bm{\Pi_f}|}{|\bm{\Pi_i}|} e^{-\frac{{b_0}^2}{2}\bm{q_\perp}^2-\frac{{b_1}^2}{2}{q_z^2}}Tr[\bar{R}_{fi,n}(p_f+mc)R_{fi,n}(p_i+mc)]\nonumber\\
\end{eqnarray}

Where:
\begin{eqnarray}
R_{fi,n}&=&\sum_s
M_{-n-s}(\Pi_f,\Pi,\rlap{$\slash$}\epsilon_c,\eta^1_{\Pi,\Pi_f},\eta^2_{\Pi,\Pi_f})
\frac{i}{\rlap{$\slash$}p-mc}
\bar{M}_{-s}(\Pi_i,\Pi,\gamma^0,\eta^1_{\Pi,\Pi_i},\eta^2_{\Pi,\Pi_i})\nonumber\\
&&+\sum_{s^\prime}M_{-n-s^\prime}(\Pi_f,\Pi^\prime,\gamma^0,\eta^1_{\Pi^\prime,\Pi_f},\eta^2_{\Pi^\prime,\Pi_f})
\frac{i}{\rlap{$\slash$}p^\prime-mc}
\bar{M}_{-s^\prime}(\Pi_i,\Pi^\prime,\rlap{$\slash$}\epsilon_c,\eta^1_{\Pi^\prime,\Pi_i},\eta^2_{\Pi^\prime,\Pi_i})\nonumber\\
\end{eqnarray}

With the argument is defined as:
\begin{eqnarray}
\eta^1_{p_1,p_2}&=&\frac{eA_0}{\hbar
c}\delta[\frac{{p_2}\cdot{\epsilon_1}}{k\cdot
p_{2}}-\frac{{p_1}\cdot{\epsilon_1}}{k\cdot p_1}]\nonumber\\
\eta^2_{p_1,p_2}&=&-\frac{eA_0}{\hbar
c}(1-\delta^2)^{1/2}[\frac{{p_2}\cdot{\epsilon_1}}{k\cdot
p_{2}}-\frac{{p_1}\cdot{\epsilon_1}}{k\cdot p_1}]\nonumber\\
\end{eqnarray}

$\Pi,\Pi^\prime$ is the four momentum of the intermediate electron
in the Feynman diagrams shown in Fig.2. They are determined by the
conservation law, together with the four momentum q transfer from
the pondermotive potential, which is given by the corresponding
functions during the calculation process:

\begin{eqnarray}
\Pi&=&\pi_f-(n+s)\hbar k+\hbar k^\prime\nonumber\\
\Pi^\prime&=&\pi_i-s\hbar k-\hbar k^\prime\nonumber\\
\hbar q&=&\pi_f-\pi_i+\hbar k^\prime-n\hbar k\nonumber\\
\end{eqnarray}

M is a $4\times4$ matrix with five arguments:
\begin{eqnarray}
M_{s}(p_1,p_2,F,\eta^1_{p_1,p_2},\eta^2_{p_1,p_2})&=&\big[\rlap{$\slash$}F+\frac{e^2{A_0}^2}{8c^2}\frac{\rlap{$\slash$}k\rlap{$\slash$}F\rlap{$\slash$}k}{(p_i\cdot
k)(p_2\cdot k)}\big]G^0_s(\alpha,\beta,\varphi)\nonumber\\
&&+\frac{eA_0}{2c}\delta\big[\frac{\rlap{$\slash$}\epsilon_1\rlap{$\slash$}k\rlap{$\slash$}F}{(p_1\cdot
k)}+\frac{\rlap{$\slash$}F\rlap{$\slash$}k\rlap{$\slash$}\epsilon_1}{(p_2\cdot
k)}\big]G^1_s(\alpha,\beta,\varphi)\nonumber\\
&&+\frac{eA_0}{2c}(1-\delta^2)^{1/2}\big[\frac{\rlap{$\slash$}\epsilon_2\rlap{$\slash$}k\rlap{$\slash$}F}{(p_1\cdot
k)}+\frac{\rlap{$\slash$}F\rlap{$\slash$}k\rlap{$\slash$}\epsilon_2}{(p_2\cdot
k)}\big]G^2_s(\alpha,\beta,\varphi)\nonumber\\
&&+(\delta^2-\frac{1}{2})\frac{e^2{A_0}^2}{4c^2}\frac{\rlap{$\slash$}k\rlap{$\slash$}F\rlap{$\slash$}k}{(p_i\cdot
k)(p_2\cdot k)}G^3_s(\alpha,\beta,\varphi)\nonumber\\
\end{eqnarray}

The generalized Bessel functions are given by:
\begin{eqnarray}
G^0_s(\alpha,\beta,\varphi)&=&\sum_n
{J_{2n-s}(\alpha)J_n(\beta)e^{i(s-2n)\varphi}},\nonumber\\
G^1_s(\alpha,\beta,\varphi)&=&\frac{1}{2}\big(G_{s+1}^0(\alpha,\beta,\varphi)+G_{s-1}^0(\alpha,\beta,\varphi)\big),\nonumber\\
G^2_s(\alpha,\beta,\varphi)&=&\frac{1}{2i}\big(G_{s+1}^0(\alpha,\beta,\varphi)-G_{s-1}^0(\alpha,\beta,\varphi)\big),\nonumber\\
G^3_s(\alpha,\beta,\varphi)&=&\frac{1}{2}\big(G_{s+2}^0(\alpha,\beta,\varphi)+G_{s-2}^0(\alpha,\beta,\varphi)\big).\nonumber\\
\end{eqnarray}

With the corresponding argument:
\begin{eqnarray}
\alpha&=&[(\eta^1_{p_1,p_2})^2+(\eta^2_{p_1,p_2})^2]^{1/2}\nonumber\\
\beta&=&\frac{Qm^2c^2}{8\hbar}(2\delta-1)\big(\frac{1}{k\cdot
p_1}-\frac{1}{k\cdot p_2}\big)\nonumber\\
\varphi&=& arctan(-\frac{\eta^2_{p_1,p_2}}{\eta^1_{p_1,p_2}})\nonumber\\
\end{eqnarray}

The above differential cross-section is related to the spontaneous
photon in the frequency interval $d\omega^{\prime}$ within the solid
angle $\Omega^{\prime}$ and the final electron within the solid
angle $\Omega_f$ . But it's difficult to detect the photon and
electron in the same time during an actual experiment. So we
integrate the differential cross-section over the scattering
electron direction, by which leads to the doubly differential cross
section only differential in the direction of the radiated photon
and its frequency:

\begin{eqnarray}
\frac{d\sigma}{d\omega^\prime d\Omega^\prime}=\int
\frac{d\stackrel{\sim}{\sigma}}{d\omega^\prime d\Omega^\prime
d\Omega_f}d\Omega_f
\end{eqnarray}

As one of the characteristic feature of a second order process, the
resonance will happen when the intermediate electron fall within the
mass shell. The physical interpretation, according to
Roshchupkin \cite{30}, may be that the considered second order
process effectively reduces to two sequential first order processes
under certain resonance condition. In the paper, the two sequential
lower processes are nonlinear Compton scattering and static
ponderomotive potential scattering, respectively. The corresponding
resonance condition may be written as $\Pi^2={m^\prime}^2c^2$ or
${\Pi^\prime}^2={m^\prime}^2c^2$. When resonance occurs, the
scattering cross-section will be divergent, which indicates that the
perturbation method is not applicable in such situation. To avoid
the problem, one has to introduce a small imaginary part of the mass
which results from the high radiative correction, i.e, the
self-energy of the laser-dressed electron, which is determined by
the total probability of the Compton scattering in a laser wave
$W_c(k\cdot p)$: $\Gamma_m(k\cdot p)=\frac{\Pi^0}{2m}W_c(k\cdot p)$
. Then the shifted propagator reads:

\begin{eqnarray}
\frac{1}{p-mc}=\frac{\Pi+m^{\prime}c}{\Pi^2-{m^\prime}^2c^2-i\frac{m}{\Pi_f^0c}(\hbar\omega^{\prime}-(n+s)\hbar\omega_0)\Gamma_m(k\cdot
\Pi_f)+im\Gamma_m(\hbar k\cdot k^{\prime})}
\end{eqnarray}

\begin{eqnarray}
\frac{1}{p^{\prime}-mc}=\frac{\Pi^{\prime}+m^{\prime}c}{{\Pi^\prime}^2-{m^\prime}^2c^2+i\frac{m}{\Pi_f^0c}(\hbar\omega^{\prime}-s\hbar\omega_0)\Gamma_m(k\cdot
\Pi_i)-im\Gamma_m(\hbar k\cdot k^{\prime})}
\end{eqnarray}

The inclusion of the imaginary part of the electron mass eliminates
the resonance singularity, which enable us to evaluate the cross
section numerically. Then, the resonance peak altitude is determined
by the lifetime of the immediate electron. Actually, the calculation
of $\Gamma_m$ has been discussed in many papers and we can easily
obtain the imaginary mass by taking advantage of the corresponding
results of our previous study.
%

\section*{3. Numerical Results}

\begin{figure}
\includegraphics[scale=0.5]{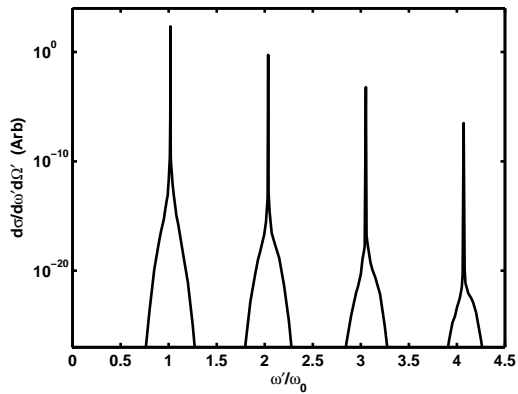}
\caption{\footnotesize{ The cross section for the low order
harmonics at emission angle $\theta^\prime=1^\circ$. Here we
consider an electron with initial energy 5MeV collide with a
circularly polarized laser with intensity parameter $Q=17.8$ and is
scattered by an effective ponderomotive potential for a $90^\circ$
geometry. The ponderomotive potential is characterized by two
parameter: $b_0=5\mu m$, $b_1=100\mu m$, describing the beam waist
size and duration of the tightly focused laser pulse,
respectively.}}
\end{figure}
%

In this section, we will present the numerical results of the
differential cross section referring to the case of $90^\circ$
laser-electron interaction geometry shown in Fig 1. We consider the
plane wave laser to be circularly polarized with a frequency of
$\omega=1.17eV$ and the dimensionless parameter Q=17.8, which is
related to a laser intensity of $7.58\times10^{22}W/cm^2$. It should
be emphasized that the numerical results in our presentation are
more exploratory than systematic since we shall focus on the
influence of ponderomotive potential due to the intensity gradients
on the photon radiation. First, we start with the results for a
tightly focused laser pulse with the parameters $b_0=5\mu m$,
$b_1=100\mu m$, which means the pulse duration is very long (i.e.,
for a pulse with wavelength $0.1\mu m$ , the duration is about
0.3ps). The corresponding differential cross section of the radiated
photon for an emission angle $\theta^{\prime}=1^\circ$ is shown in
Fig 3. As expected, high harmonics are generated and the positions
of the resonances located in the spectrum coincide with those for
NLCS. This is due to the fact that the resonant second order process
can effectively reduce to two sequent lower order processes as
mentioned before while the mechanism responsible for radiation of
photons is the nonlinear Compton scattering process. Furthermore,
the main contribution of the cross section comes from
$\bm{q}\approx0$ at a resonance, which means there's nearly no
momentum transfer between the electron and the ponderomotive
potential. It's easy to find that the main influence due to the
ponderomotive potential is the broadened spectrum compared to that
of NLCS, which is in accordance with the conclusion based on a
classical theory \cite{24}. The magnitude of the spectrum drops fast
when the energy of photon is away from resonance. From a
mathematical point of view, that's because the differential cross
section is subject to an exponential decay with respect to the
transfer momentum q. So the spectrum broadening effects is
determined both the parameter $b_0$ and $b_1$, or the pulse duration
and the beam waist size.

\begin{figure}
\includegraphics[scale=0.5]{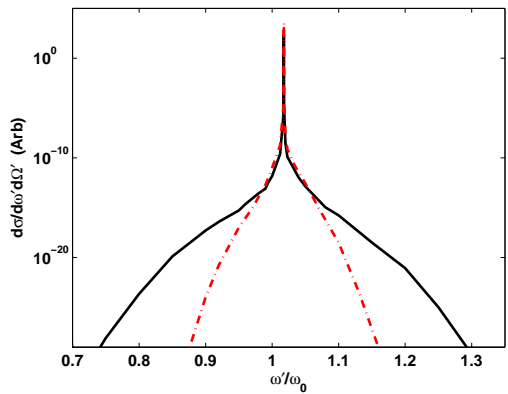}
\caption{\footnotesize{The cross section for the fundamental
harmonic at emission angle $\theta^\prime=1^\circ$. The condition is
same with Fig 3 except the parameter $b_0$: $5\mu m$ for full line
and $10\mu m$ for dot-dashed line.}}
\end{figure}
%

Now we proceed to investigate how the broadening effect depends on
the laser pulse duration and the beam waist size. In Fig 4, we
compare the spectrum of fundamental harmonics for different beam
waist size with same pulse duration. The cross section with
$b_0=5\mu m$ is denoted by full line, whereas the dotted-dashed line
denote the cross section with $b_0=10\mu m$. Here the magnitude of
the cross section near resonance is a little larger for a broader
beam waist size. This is probably because a laser with certain
intensity has a great energy with the increase of the beam waist
size. We can also see from the expression (11) that the differential
cross-section is proportional to the quartic of $b_0$. But the peak
falls off much more quickly for a larger beam waist size. This can
be understood by considering that the differential cross-section is
determined exponentially by $b_0$. The physical interpretation may
be that the large beam waist leads to a less field-gradient, which
means a small ponderomotive potential effect. When the beam waist is
so large that we can neglect the spatial gradient if we consider the
interaction of the electron with laser field near its focus place,
there will be almost no ponderomotive broadening at all. In the
following,we shall refer to the spectra of different laser pulse by
changing the parameter $b_1$, as shown in Fig 5. The difference is
so small that it can be hardly visible. We thus find that the pulse
duration plays only a minor role in the spectrum. That's probably
because there's far less momentum transfer from the electron onto
the ponderomotive potential in the laser propagation direction
(i.e., $q_z^2\approx0\ll q_\perp^2$) for a $90^\circ$ interaction
geometry. Hence, there is almost no ponderomotive potential
scattering in this direction.

\begin{figure}
\includegraphics[scale=0.5]{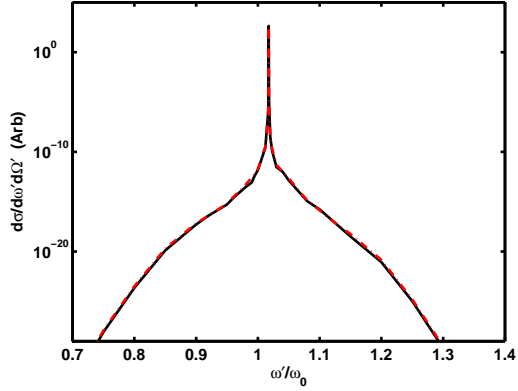}
\caption{\footnotesize{The cross section for the fundamental
harmonic at emission angle $\theta^\prime=1^\circ$. The condition is
same with Fig 3 except the parameter $b_1$: $b_1=100\mu m$ for full
line and $b_1=150\mu m$ for dot-dashed line.}}
\end{figure}
%

Finally, we compare the results of different laser intensity of the
circularly polarized plane wave field. The corresponding spectra are
displayed in Fig 6. It shows that the radiation for the plane wave
field Q=5 (corresponds to a intensity of $3\times10^{19}W/cm^2$) is
several magnitudes smaller than that for Q=17.8 (corresponds to a
intensity of $7.58\times10^{20}W/cm^2$). But the broadening width of
the spectra is similar for two different laser strengths. It
confirmed the conclusion that the broadening effect has almost
nothing to do with the plane wave field, but caused by the
ponderomotive potential

\begin{figure}
\includegraphics[scale=0.5]{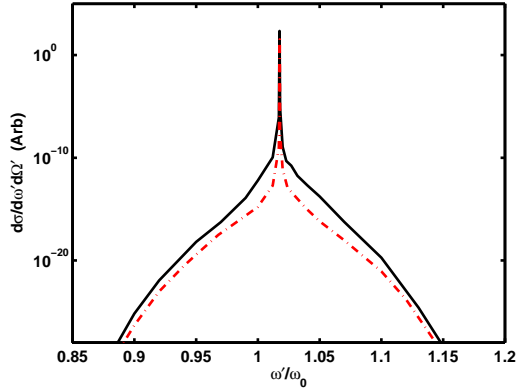}
\caption{\footnotesize{The cross section for the fundamental
harmonic at emission angle $\theta^\prime=1^\circ$. The condition is
same with Fig3 except the intensity of the plane wave laser:
Q=17.8($7.58\times10^{20}W/cm^2$) for full line and
Q=5($3\times10^{19}W/cm^2$) for dot-dashed line.}}
\end{figure}
%


\section*{4. Summary}
In this paper, we study the role of the field intensity gradients on
the radiation spectrum emitting from the electron in the laser
field. The whole scattering process is calculated as a
laser-modified second order QED process with resonances addressed.
Consequently, it shows that the spectrum is broadening due to the
ponderomotive effects compared to that of NLCS, with the positions
of the resonance peak corresponding to the discrete harmonic
frequencies of NLCS. Because the broadening of the spectrum line contains important information about the laser intensity gradient, it might provide
us a feasible way to measure the width of the intense laser beam. In addition, the broadening effect is
determined far more by the beam waist size than the pulse duration
for the case of $90^\circ$ incidence scattering geometry.
Since the parameters of the corresponding laser field is
readily accessible in the lab, it is hoped that these results could be submitted to the
experiment test in the near future.
\acknowledgements This work is supported by the National Natural Science
Foundation of China under Grant Nos. 10974056 and 11274117. One of the authors, Wenjun Zhu, thanks the support by the Science and Technology Foundation of National Key Laboratory of Shock Wave and Detonation Physics (Grant No. 077110).\\


\end{document}